# Sign reversal of field like spin-orbit torque in ultrathin Chromium/Nickel bilayer

Arnab Bose*, Hanuman Singh, Swapnil Bhuktare, Sutapa Dutta and Ashwin A. Tulapurkar

*Dept. of Electrical Engineering, Indian Institute of Technology Bombay, India-400076*
*arnabbose@ee.iitb.ac.in

**Abstract**

Relativistically originated spin-orbit torque is one of the promising ways to control magnetization dynamics of ferromagnet which can be useful for next generation spintronic memory applications. Lot of effort has been made to address the physical origin of spin-orbit torque and improve its efficiency. In this work we demonstrate that in ultrathin chromium /Nickel (Cr/Ni) hetero-structure, spin-orbit torque significantly increases for Cr thickness below 6 nm. We have also observed unconventional sign of field like torque which can be attributed to the interfacial Rashba kind of coupling. We experimentally obtain that approximately 35 Oe Rashba kind of magnetic field is created on 8 nm thick in-plane magnetized Ni film when $10^8 A/cm^2$ current density flows through Cr layer.

**Introduction**

Since last few years relativistic spin-orbit torque [1] has been central focus among the researchers to manipulate nanomagnets. Spin orbit torque can arise from pure spin current, generated from bulk spin Hall effect [2,3,4]. It can also be originated from effective magnetic field due to interfacial Rashba [5-6] or Dresselhous [7] spin orbit interaction where structural or crystal inversion symmetry is broken. Extensive research on spin-orbit torque has been done on different materials including heavy metals (HM) like Pt [8-9], Ta [10-11], W [12,13], antiferromagnets like PtMn, IrMn, FeMn [14,15,16,17], topological insulators [18,19,20], 2D materials like $WTe_2$ [21], $WSe_2$ [22] and semiconductors [23,24,25,26] . Spin-orbit torque has applications in nano rf-oscillator [27,28,29] and magnetic switching [10,14,30,31]. For muli-layered metallic systems where structural inversion symmetry is broken at the interface, spin orbit torque can arise from both Rashba interaction and bulk spin Hall effect. Generally it is difficult to identify Rashba spin orbit torque since its action is same as Oersted field. For example, switching of Co in $Pt/Co/AlO_x$ heterostructure is ascribed to bulk SHE [30] and Rashba field [31]. In this work we experimentally demonstrate that spin-orbit torque is significantly enhanced in Chromium/Nickel (Cr/Ni) bilayer when Cr thickness is below 6 nm. Most interestingly we report unprecedented observation of net field like torque [32] showing opposite sign compared to Oersted field torque. This sign reversal of field like torque could be ascribed to interfacial Rashba coupling. To confirm the interfacial nature of the torque, we inserted Cu interlayer between Ni and Cr and found that the sign of field like torque returns to be same as Oersted field in Cr/Cu/Ni unlike Cr/Ni bilayer.

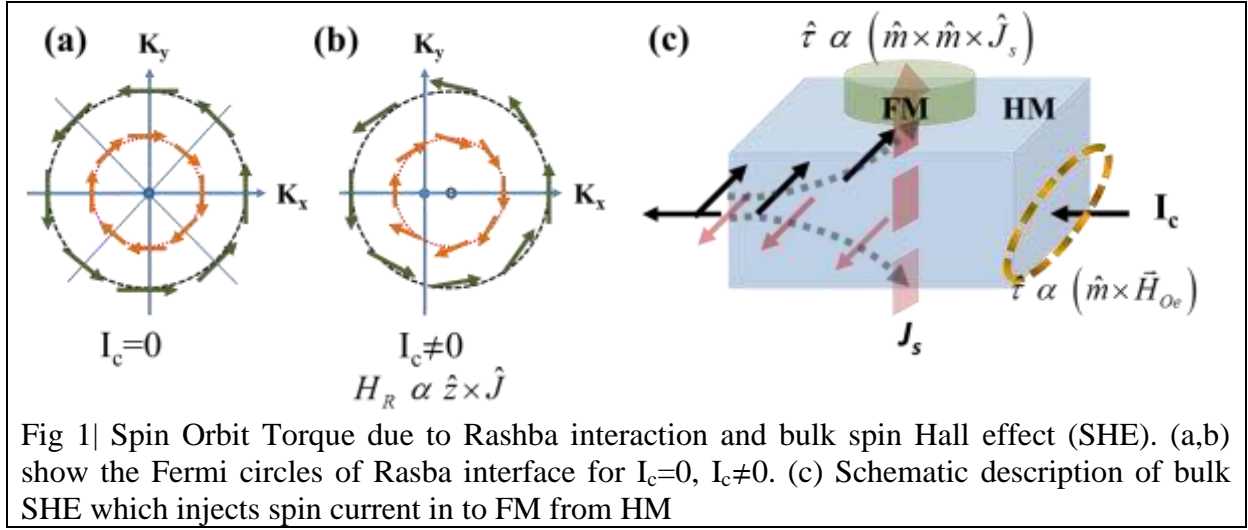

Fig 1| Spin Orbit Torque due to Rashba interaction and bulk spin Hall effect (SHE). (a,b) show the Fermi circles of Rasba interface for $I_c=0$, $I_c\neq 0$. (c) Schematic description of bulk SHE which injects spin current in to FM from HM

Figure 1.a,b shows Fermi circles corresponding to the two different chiral spin states of two dimensional Rashba interface. Without any applied current Fermi circles are symmetric in momentum space leading to zero spin accumulation (Fig 1.a). When electric current is passed through Rashba interface it breaks the symmetry in momentum space which causes imbalance of accumulated spins (Fig 1.b). Since Fermi radius is different for two spin states, we get a net spin accumulation perpendicular to the direction of current flow (y direction in this case.) These spins at the interface interact with spins of ferromagnet (which is on top of HM) by exchange interaction. It leads to the generation of effective magnetic field perpendicular to charge current direction which can be described by following expression $\bar{H}_{eff,R} = \alpha_R (\hat{z} \times \bar{J}_C)$ where $\alpha_R$ is coupling constant which depends on material, z is out of plane direction along which electric field is created due to structural inversion asymmetry and $J_C$ is the charge current density. This magnetic field ($H_{eff,R}$) has same symmetry like Oersted field ($H_{Oe}$) but in our experiment it ($H_{eff,R}$) completely dominates over $H_{Oe}$ changing the sign of field like torque which has following expression $\vec{\tau}_F \alpha (\hat{m} \times \vec{H})$ where m is unit vector along magnetization and H is magnetic field. In the experiment we also observed the contribution to spin orbit torque from bulk SHE [$\bar{\tau}_{STT} \propto (\hat{m} \times \hat{m} \times \bar{I}_S)$], where the spin current, $I_s$ is given by $\bar{I}_S \propto \theta_{sh} (\hat{z} \times \bar{J}_C)$. The vector sign on spin current ($I_s$) denotes the direction of spin polarization.

**Experiment**

Two well-known methods are largely adopted by researchers to quantify spin-orbit torque viz. (1) spin orbit torque ferromagnetic resonance (ST-FMR) [8,15,16,23,24] and (2) harmonic measurements [11,33]. We have adopted the first one: STFMR which occurs due to spin torque diode effect [34,23,8]. Schematics of the experimental set up is shown in the figure 2.a. Radio frequency (RF) current is passed through HM/Ni bilayer (HM is either Cr or Pt) while in-plane magnetic field is applied at angle θ with respect to current flow direction. Resultant DC voltage is measured using bias-T network sweeping external magnetic field. RF current passing through HM is converted into pure spin current by SHE. This spin current is absorbed by ferromagnetic Ni film. RF current in HM also produces both Oersted field (from bulk) and effective Rashba magnetic field (from the interface). Both these fields ($H_{eff,R}$ and $H_{Oe}$) and injected spin current exert torque on Ni which causes small oscillation of its magnetization.

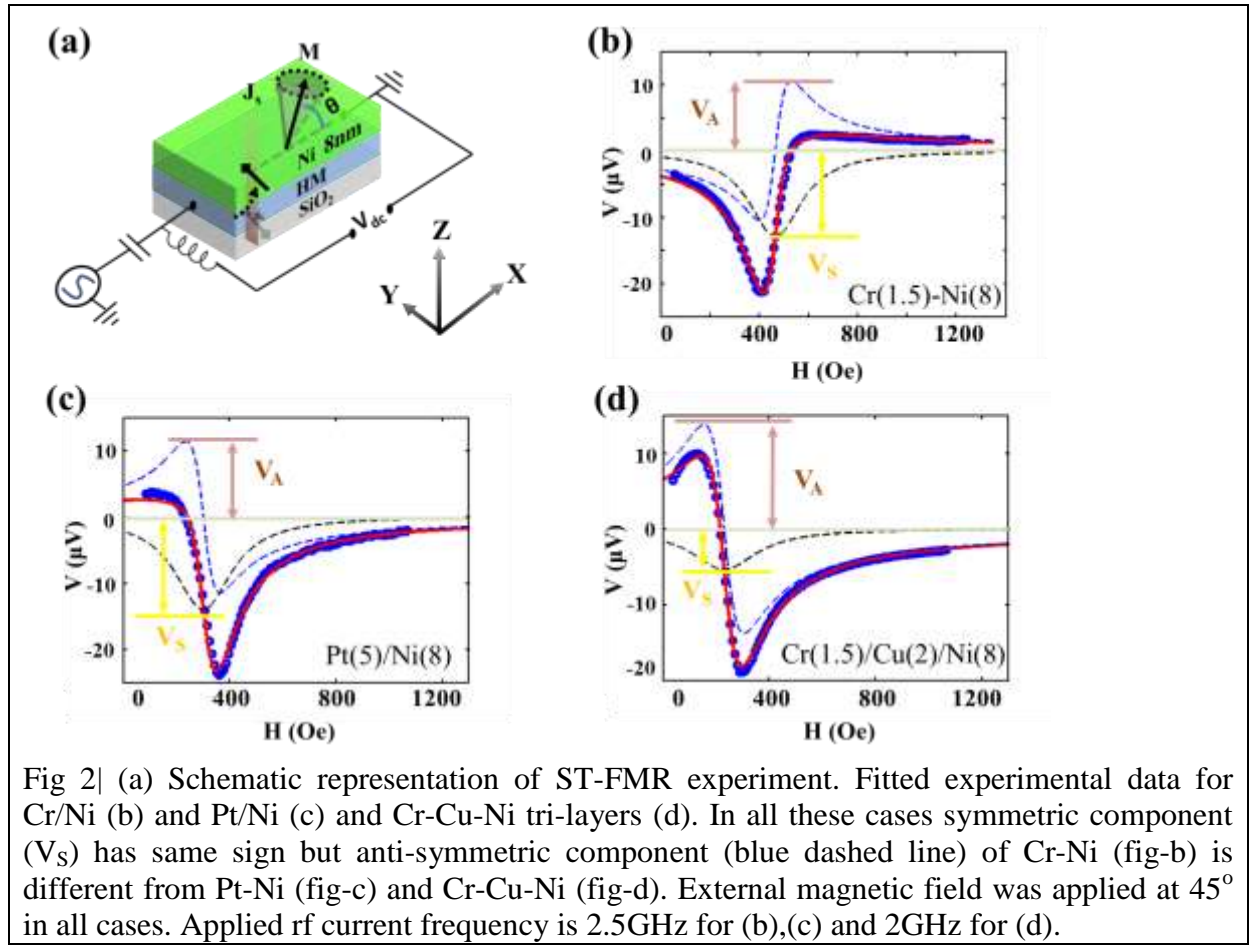

Fig 2| (a) Schematic representation of ST-FMR experiment. Fitted experimental data for Cr/Ni (b) and Pt/Ni (c) and Cr-Cu-Ni tri-layers (d). In all these cases symmetric component ($V_S$) has same sign but anti-symmetric component (blue dashed line) of Cr-Ni (fig-b) is different from Pt-Ni (fig-c) and Cr-Cu-Ni (fig-d). External magnetic field was applied at 45° in all cases. Applied rf current frequency is 2.5GHz for (b),(c) and 2GHz for (d).

As magnetization oscillates, the resistance of Ni also oscillates due to the anisotropic magnetoresistance (AMR) effect. The homodyne mixture of oscillatory resistance and RF current produces measurable DC voltage at resonance. All experiments are carried out at room temperature, in three different types of samples: Pt(6nm)/Ni(8nm), Cr(t)/Ni(8nm) and Cr(1.5nm)/Cu(2nm) /Ni(8nm). Thickness of Cr in second sample was varied from t=1.5 to 8 nm. Figure 2.b-d show typical behaviour of spin orbit torque in HM/FM bilayer. It is well fitted to following symmetric Lorentzian ($V_s$) and anti-symmetric ($V_A$) dispersion relations:

$V_S = C_1 \dfrac{\Delta^2}{4(H-H_0)^2 + \Delta^2}$ and $V_A = C_2 \dfrac{4(H-H_0)\Delta}{4(H-H_0)^2 + \Delta^2}$ where $H_0$ is resonance field and $\Delta$ is linewidth. $C_1$ represents the strength of symmetric Lorentzian ($V_S$) which corresponds to the in-plane spin orbit torque due to injected spin current from Cr (or Pt) to Ni generated by SHE and $C_2$ represents the strength of anti-symmetric component, corresponds to the in-plane effective field originated from combination of Oersted field and interfacial effective Rashba field. Comparing figure 2.b and 2.c we can clearly say that symmetric component of Pt/Ni and Cr/Ni have same sign but anti-symmetric component has different sign. This is very unique observation we are reporting in Cr/Ni bilayer compared to other (metallic) systems studied so far. We have verified that sign of field like torque in Pt/Ni bilayer is same as Oersted field which other groups also reported [8,15,16,23,24]. Hence our observation in Cr/Ni bilayer cannot be explained by Oersted field. To find the origin of such unconventional behaviour of field like torque, 2 nm thick Cu is inserted between Cr and Ni where spin transport happens through diffusive regime of Cu spacer. We can clearly see that on insertion of Cu between Cr and Ni, sign of field like torque (FLT) shows same behaviour of Oersted field [Fig 2.d]. So it is confirmed that interface played the pivotal role to invert the sign of FLT which can be explained by generation of Rashba kind of field at Cr/Ni interface. We have systematically studied thickness and angular dependence of spin orbit torque in Cr/Ni bilayer which are shown in figure 3. We observe that with decrease of Cr thickness its resistivity increases expectedly [35,36] (Fig 3.d) and ST-FMR signal also improves (Fig 3.a). We have done comparative study of angular dependence of spin orbit torque for both Cr/Ni and Pt/Ni heterostructure devices which show expected $\sin 2\theta \cos\theta$ dependence (Fig 3) (except the sign change of FLT). Even for different angle ($\theta$) and thickness of Cr ($t_{Cr}$: 1.5, 2.5, 4, 5, 6, 8 nm) field like torque of Cr(t)/Ni(8nm) bilayer shows opposite behaviour compared to Pt/Ni (Fig 3.a,c,f). This consistent result cannot be explained by any non-uniform current distribution through Ni and few oxidized monolayers (1nm to 2 nm) on top of Ni surface.

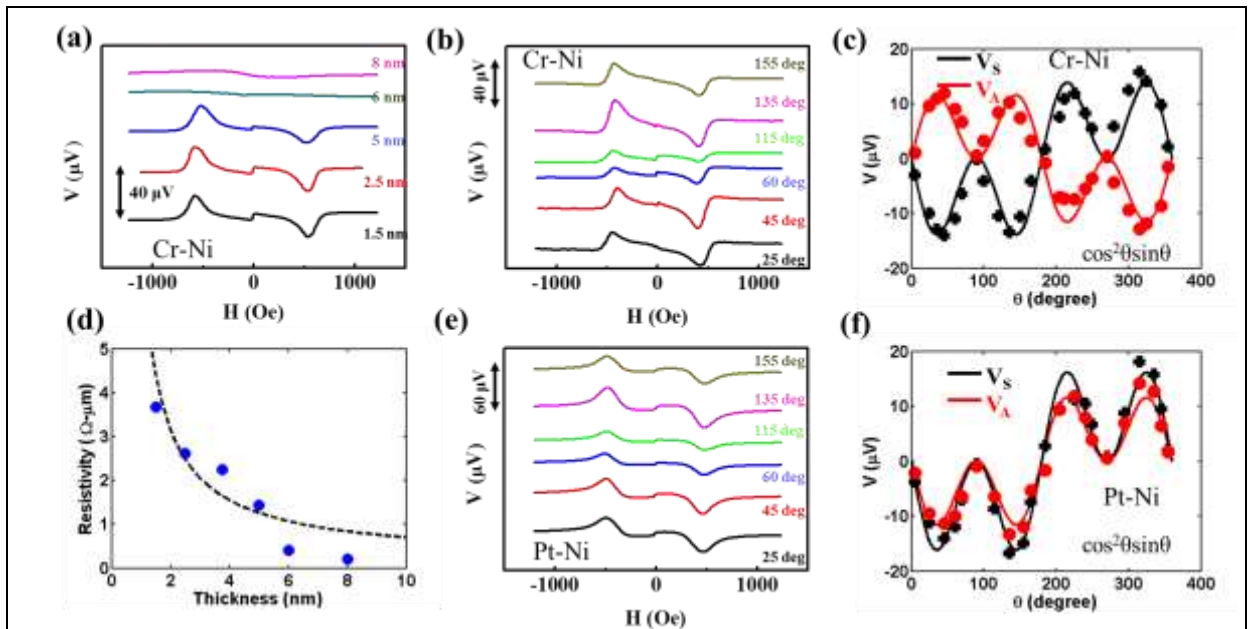

Fig 3| Study of angular dependence and thickness dependence. (a)ST-FMR spectrum for different thicknesses of Cr in Cr(t)/Ni(8) bilayer, for external magnetic field applied at 45° angle. (b),(c) Angular dependence of ST-FMR signal of Cr(1.5)/Ni(8) bilayer. (d) Resistivity of Cr as a function of its thickness. (e), (f) Angular dependence of ST-FMR signal of Pt(5)/Ni(8) bilayer. Frequency of applied RF current is 2.5 GHz.

**Discussion**

In this section we quantify the contribution of bulk spin Hall effect and interfacial Rashba originated magnetic field. We solved below LLG equation for small oscillation of magnetization around equilibrium state:

$$\dot{\hat{m}} = -\gamma_0 \, \hat{m} \times (\bar{H}_{eff} + \bar{h}_{ac}) + \alpha(\hat{m} \times \dot{\hat{m}}) + A_I(\hat{m} \times \bar{I}_s \times \hat{m})$$

Here γ denotes gyromagnetic ratio, α denotes damping constant, $H_{eff}$ is effective magnetic field comprising of external field and anisotropy field, $h_{ac}$ denotes magnetic field arising from Oersted magnetic field and Rashba field. The coefficient $A_I$ is defined as $A_I = \mu_B/qM_sV$, where $\mu_B$ denotes Bohr magneton, $M_s$ denotes saturation magnetization and V denotes volume of FM. The spin current density incident on FM is given by, $\bar{J}_S = \theta_{sh} J_c \hat{y}$, where $\theta_{sh}$ is spin Hall angle and $J_c$ is charge current density in HM flowing along x direction. Rashba magnetic field is given by, $\bar{H}_{eff,R} = \alpha_R J_C \hat{y}$ where $\alpha_R$ denotes Rashba coefficient. DC voltage in the sample is given by, $V_{dc} = -\Delta R \cos^2\theta \sin\theta \, I_{ac}[real(\chi_{11}^H)h_{ac} + real(\chi_{11}^{sc})(A_I/\gamma)I_S]$, where θ denotes the angle between the equilibrium magnetization direction and x-axis (direction of current flow), $\chi_{11}^H$ and $\chi_{11}^{sc}$ denote the susceptibilities to magnetic field and spin current respectively. The real part of these susceptibilities has dispersion and Lorenzian functions respectively. The resistance of the sample is assumed to vary as, $R = R_P - \Delta R(1-m_x^2)$ due to the AMR effect. We have used following parameters in the simulation: AMR 0.5%, applied RF power 10dBm, damping constant 0.07, out-of plane anisotropic field 2.5 kOe. We observe large enhancement of spin Hall angle (or charge to spin current conversion efficiency) while thickness of Cr is decreased (left axis of fig 4). After fitting we obtain that, for 1.5 nm of thickness of Cr spin Hall angle becomes +0.085 which is very much comparable to other reported value in case of Pt. However in case of our Pt/Ni bilayer spin Hall angle turns out to be +0.15 slightly higher compared to other reports which can be attributed to impurity present in Pt for different condition. This trend of increasing spin Hall angle with decreasing thickness of heavy metals is also recently reported in YIG/Au system [37]. In contrast to SHE, effective interfacial magnetic field generated by Rashba spin orbit interaction ($H_{eff-R}$) is approximately 35 Oe for $10^8$ A/cm$^2$ current density through Cr layer and it is experimentally observed to be almost independent of Cr thickness (right axis of fig 4). $H_{eff-R}$ in our experiment is smaller compared to previous report [9] where thickness of FM was ten times smaller compared to this work. However our experimental result differs to some of the other previous reports. Recently inverse spin Hall effect (ISHE) is reported in Cr by spin pumping method [38] and by longitudinal spin Seebeck method [39] with opposite sign of spin Hall angle of Cr with respect to Pt. We surprisingly observed that in our experiment, spin Hall angle of Cr has same sign of Pt and it has quite large value (+0.085) for lower thickness of Cr (Fig 2.b,c). Unlike previous reports of ISHE in Cr [38,39], we do not observe spin orbit torque for higher thickness of Cr ($t_{Cr}$>6 nm). One possible reason could be the extrinsic nature of spin Hall effect. We have neglected any sort of voltage contribution from ISHE and Inverse Rashba Edelstein effect (IREE) due to spin pumping of Ni since it is order of magnitude smaller compared to ST-FMR [8]. Very recently it is predicted that in spin valve structure with current perpendicular to plane geometry, field like torque can also change its sign if strong exchange interaction exists between itinerant electrons and localized magnetization [40]. Our work presents the geometry where current flows in plane through a heavy metal and ferromagnet. s orbital and d orbital electron coupling in Ni is not strong enough to explain our results: reversal of field like torque in Cr/Ni but not in Pt/Ni or Cr/Cu/Ni. Neel temperature of bulk Cr is 311K and for thin film it is below room temperature [38,39]. All our experiments are carried out at room temperature, and we can assume Cr to be paramagnetic. From our experiment it is evident that interfacial Rashba originated spin orbit torque could be one of the possible reasons for sign reversal of field like torque. However detailed theoretical study and more experimental investigation are required to explore the complete picture of this unusual behaviour of spin torque.

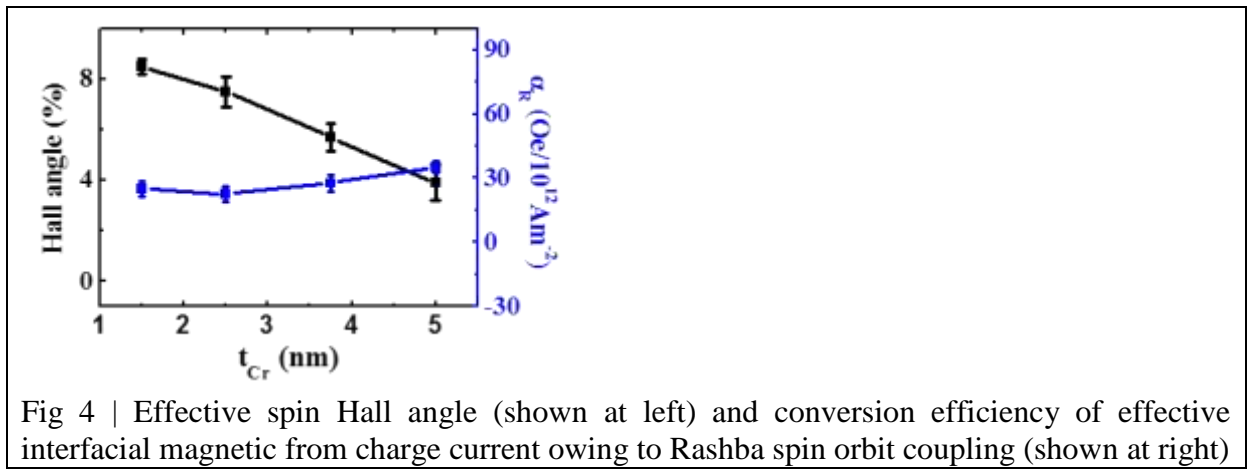

Fig 4 | Effective spin Hall angle (shown at left) and conversion efficiency of effective interfacial magnetic from charge current owing to Rashba spin orbit coupling (shown at right)

**Conclusion**

In conclusion we found that lower thickness of Cr (<6 nm) shows high spin orbit torque which can arise from interfacial Rashba spin-orbit interaction and bulk spin Hall effect. Unconventional sign change in field like torque is observed in Cr/Ni bilayer unlike other Cr/Cu/Ni and Pt/Ni heterostructures. It certainly assures the origin of sign reversal is interfacial which can be attributed to Rashba like spin-orbit torque. The high efficiency of these combined spin orbit torques (spin Hall effect and Rashba magnetic field) can be useful for future memory application.

**Acknowledgement**

We are thankful to the Centre of Excellence in Nanoelectronics (CEN) at the IIT-Bombay Nanofabrication facility (IITBNF) and Department of Electronics and Information Technology (DeitY), Government of India for its support. Athors thank Pijush Chakroborty for helping in sputter deposition.